\documentclass[aps,showpacsx,twocolumn]{revtex4-1}
\usepackage{graphicx}
\usepackage{bm}
\usepackage{slashbox}
\usepackage{color}
\usepackage{eufrak}
\usepackage{mathrsfs}

\begin{document}

\title{Critical Behavior and Dimension Crossover of Pion Superfluidity}
\author{Ziyue Wang and Pengfei Zhuang}
\affiliation{Physics Department, Tsinghua University and Collaborative Innovation Center of Quantum Matter, Beijing 100084, China}
\date{\today}
\begin{abstract}
We investigate the critical behavior of pion superfluidity in the frame of functional renormalization group. By solving the flow equation in the SU(2) linear sigma model with pion superfluidity at finite temperature and isospin density, and making comparison with the fixed point analysis of the flow diagram in a general $O(N)$ model at zero temperature and density but with continuous dimension, the pion superfluidity is a second order phase transition subject to a $O(2)$ universality class with a dimension crossover from $d_{\text{eff}}=4$ to $d_{\text{eff}}=3$. The usually used large $N$ expansion gives a temperature independent critical exponent $\beta$ and agrees with the renormalization group only at zero temperature.
\end{abstract}
\maketitle

\section{Introduction}
The study of Quantum Chromodynamics (QCD) phase transitions at finite temperature and density provides a deep insight into the strong interacting matters created in high energy nuclear collisions and compact stars. The QCD symmetry breaking and restoration patterns lead to a very rich structure in the QCD phase diagram. The extension of the phase diagram from finite temperature and baryon density to finite isospin density is motivated by the investigation of isospin imbalance in the interior of neutron stars~\cite{Barshay:1974,Khodel:2004}. The thermodynamic equilibrium systems with finite isospin chemical potential have been widely investigated through lattice simulations~\cite{Endrodi:2014, Kogut:2004, Kogut:2002} and effective models like the Nambu--Jona-lasinio (NJL) model~\cite{He:2005,Andersen:2007,Abuki:2008,Mu:2009,Xia:2013}, quark-meson model~\cite{Kamikado:2012} and linear sigma model~\cite{Svanes:2010,Phat:2011}. It is found that the spontaneous breaking and restoration of the symmetry between charged pions $\pi_\pm$ is a second order phase transition at both zero and finite temperature.

As a nonperturbative method, the Functional Renormalization Group (FRG) has been used to study phase transitions in various systems like cold atom gas~\cite{Floerchinger:2009,Boettcher:2012}, nucleon gas~\cite{Friman:2012,Drews:2013}, and hadron gas~\cite{Schaefer:1999,Bohr:2000,Blaizot:2006,Braun:2008,Braun:2009,Stokic:2009,Fukushima:2010,Herbst:2013}. By solving the flow
equation which connects physics at different momentum scales, the FRG shows a great power in describing the critical behavior of the phase transitions which is controlled by quantum and thermodynamic fluctuations. As commonly concerned in continuous phase transitions, critical exponents and university class play an extremely important role in the study of QCD phase transitions. Different systems with the same symmetry and dimension should share identical critical exponents and belong to the same universality class, indicating that the collective properties near a continuous phase transition are independent of the dynamical details of the system. The FRG encodes thermal and quantum fluctuations from ultraviolet to infrared through coarse graining and scale transformation, giving reliable physical predictions in the vicinity of continuous phase transitions.

The macroscopic behavior of a thermal equilibrium system is controlled by the low momentum modes and thermodynamical parameters such as temperature and chemical potential. For a system coupled to an external heat bath, the change in the degrees of freedom may lead to a change in the intrinsic symmetry. For instance, the elemental constituents of a QCD system undergo a change from hadrons at low temperature to quarks and gluons at high temperature, and this change leads to the deconfinement phase transition. On the other hand, the system goes through a continuous change in the dimension when the temperature increases. The dimension reduction phenomenon has long been investigated in effective theories~\cite{Appelquist:1981}. For a system defined in a space with continuous spatial dimension $d$ and periodic time dimension, the space-time dimension undergoes a change from $d+1$ to $d$ with increasing temperature~\cite{Liao:1996,Ballhausen:2003}.

Dimension is crucial for the critical behavior of a continuous phase transition, since it is an essential element in the classification of universality classes. The Ginzburg criterion gives the idea of an upper critical dimension~\cite{Nielsen:1977}. For a theory with space-time dimension greater than $4$, the mean field approach is already good enough to describe the critical phenomena. However, when the dimension is less than $4$ (at finite temperature), fluctuations need to be taken into account. In fact, the exact solution to the Gaussian model in space with dimension $4$ gives indeed the same result as the mean field method, despite of the difference in understanding the physics. This motivates the expansion in $4-\epsilon$~\cite{Wilson:1971,Guida:1998}, under the assumption that the critical exponents and physical parameters are continuous with the change in dimension. When the dimension is slightly less than $4$, the small deviation from the mean field result can directly be extracted from the expansion. In the frame of FRG, the scale invariance and self-similarity near the critical point are described by the scale invariant fixed point of the flow equations. The critical surface and properties of the flow around the nontrivial fixed point provide a far-reaching description concerning critical exponents and universality class.

The paper is organized as follows. The FRG application to the $SU(2)$ linear sigma model is given in Section 2, the solutions of the FRG flow equations, especially the critical exponents of the pion superfluidity, are given in Section 3, and the comparison with a $O(N)$ model in continuous dimension is discussed in Section 4. We summarize in Section 5.

%%%%%%%%%%%%%%%%%%%%%%%%%%%%%%%%%%%%%%%%
\section{The FRG Application to Pion Superfluidity}
%%%%%%%%%%%%%%%%%%%%%%%%%%%%%%%%%%%%%%%%
As an effective low energy model, the linear sigma model exhibits many of the global symmetries of QCD at meson level and is widely used to demonstrate the spontaneous chiral symmetry breaking in vacuum and its restoration at finite temperature and baryon density~\cite{Nemoto:1999,Baacke:2002,Andersen:2008,Andersen:2004}. At finite isospin chemical potential $\mu_I$, the SU(2) linear sigma model is defined through the Lagrangian density
\begin{eqnarray}
\mathcal{L} &=& {1\over 2} \partial_\mu \pi\partial^\mu\pi+i\mu_I\left(\pi_1\partial_t\pi_2-\pi_2\partial_t\pi_1\right)+U(\pi),\nonumber\\
U(\pi) &=& {m^2\over 2}\pi^2+{\lambda\over 4}\pi^4-{\mu_I^2\over 2}\left(\pi_1^2+\pi_2^2\right)-c\sigma,
\label{l1}
\end{eqnarray}
where $\pi=(\sigma, {\boldsymbol \pi})$ is defined as a 4-component field constructed by the isoscalar $\sigma$ and isovector ${\boldsymbol \pi}=(\pi_1, \pi_2, \pi_3)$. In vacuum at $\mu_I=0$, the $O(4)$ symmetry in chiral limit is explicitly broken to $O(3)$ with a rotational symmetry among the three pions in real case with $c\neq 0$. Turning on the isospin chemical potential splits the three pions, and the $O(3)$ symmetry is explicitly broken to $O(2)$ symmetry. When $\mu_I$ exceeds the pion mass $m_\pi$ in vacuum, the symmetry is further spontaneously broken from $O(2)$ to $Z(2)$ and the system enters the pion superfluidity phase.

The spontaneous chiral symmetry breaking and isospin symmetry breaking are described respectively by the chiral condensate $\langle\sigma\rangle$ and charged pion condensate $\langle\pi_+\rangle=\langle\pi_-\rangle$. Considering the relations $\pi_+=\left(\pi_1+i\pi_2\right)/\sqrt 2$ and $\pi_-=\left(\pi_1-i\pi_2\right)/\sqrt 2$, one can take alternatively the condensate $\langle\pi_1\rangle$ as the order parameter of pion superfluidity. Separating the quantum fluctuations from the classical mean fields by making a shift $\pi=(\sigma, \pi_1, \pi_2, \pi_3)\to\langle\pi\rangle+\pi=(\langle\sigma\rangle, \langle\pi_1\rangle, 0, 0)+(\sigma, \pi_1, \pi_2, \pi_3)$, the Lagrangian density becomes
\begin{eqnarray}
\mathcal{L} &=& \mathcal{L}_{mf}+\mathcal {L}_{int},\nonumber\\
\mathcal{L}_{mf} &=& {1\over 2} \partial_\mu \pi\partial^\mu\pi+i\mu_I\left(\pi_1\partial_t\pi_2-\pi_2\partial_t\pi_1\right)+{1\over 2}m_i^2\pi_i^2\nonumber\\
&& +U\left(\langle\pi\rangle\right),
\end{eqnarray}
where $U(\langle\pi\rangle)$ is the classical potential, the meson mass $m_i^2$ generated by the condensates can be expressed as the second order derivative of $U(\pi)$ with respect to the meson field at $\pi=\langle\pi\rangle$, $m_i^2=\left(\partial^2U(\pi)/\partial\pi_i^2\right)|_{\pi=\langle\pi\rangle}$, and the interaction part $\mathcal{L}_{int}$ contains 3-meson and 4-meson interactions with couplings in terms of the third and fourth order derivatives of $U(\pi)$.

The physical condensates $\langle\sigma\rangle$ and $\langle\pi_1\rangle$ are determined by the minimum potential
\begin{eqnarray}
{\partial U(\langle\pi\rangle)\over \partial \langle\sigma\rangle} &=& \langle\sigma\rangle\left(m^2+\lambda\left(\langle\sigma\rangle^2+\langle\pi_1\rangle^2\right)\right)-c=0,\nonumber\\
{\partial U(\langle\pi\rangle)\over \partial \langle\pi_1\rangle} &=& \langle\pi_1\rangle\left(m^2+\lambda\left(\langle\sigma\rangle^2+\langle\pi_1\rangle^2\right)-\mu_I^2\right)=0,
\end{eqnarray}
which guarantees the disappearance of the linear terms in $\pi$ in the interaction Lagrangian $\mathcal {L}_{int}$.

The three model parameters, namely the mass parameter $m^2$, the 4-meson coupling constant $\lambda$ and the chiral breaking parameter $c$, are fixed by fitting the meson masses $m_\pi^2=m_{\pi_1}^2=m_{\pi_2}^2=m_{\pi_3}^2=m^2+\lambda\langle\sigma\rangle^2=135$ MeV and $m_\sigma^2=m^2+3\lambda\langle\sigma\rangle^2=400$ MeV and the pion decay constant $f_\pi=\langle\sigma\rangle=93$ MeV in vacuum. To guarantee the Lorentz invariance and parity conservation, the pion condensate should vanish in vacuum, $\langle\pi_1\rangle=0$. At finite isospin chemical potential, the solution of the two coupled gap equations is $\langle\sigma\rangle = f_\pi,\ \langle\pi_1\rangle = 0$ in the normal phase at $\mu_I < m_\pi$ and $\langle\sigma\rangle = c/\mu_I^2,\ \langle\pi_1\rangle = \sqrt{(\mu_I^2-m^2)/\lambda-c^2/\mu_I^4}$ in the pion superfluidity phase at $\mu_I > m_\pi$.

The inclusion of thermal excitations in the linear model should be treated carefully. The Hartree-Fock approach is straightforward, but its disadvantage is the lack of the Goldstone mode in the symmetry breaking phase~\cite{Petropoulos:2004} at finite temperature and density. Here we introduce the thermal excitations through the large $N$ expansion~\cite{Moshe:2003} and focus on the phase diagram of pion superfluidity. Considering the $O(3)$ symmetry of the model in vacuum, we adopt the large $N$ expansion method in the $O(N)$ model with isospin chemical potential. With the same method and technics given by Harber and Weldon~\cite{Haber:1981}, the condensates $\langle\sigma\rangle$ and $\langle\pi_1\rangle$ and the related mass parameter $M^2$ are controlled by the coupled gap equations
\begin{eqnarray}
&& \langle\pi_1\rangle\left(M^2-\mu_I^2\right)=0,\nonumber\\
&& \langle\sigma\rangle M^2-c=0,\nonumber\\
&& M^2=\lambda\left(\langle\pi\rangle^2-f_\pi^2+2J'(T,\mu_I,M^2)\right)+{c\over f_\pi},
\end{eqnarray}
where the thermal excitations are included in the function $J'(T,\mu_I,M^2)$,
\begin{equation}
J' = {1\over 2}\int{d^3{\bf p}\over (2\pi)^3}{1\over E}\left(2f(E)+f(E-\mu_I)+f(E+\mu_I)\right)
\end{equation}
with the quasi-particle energy $E=\sqrt{M^2+{\bf p}^2}$ and Bose-Einstein distribution $f(x)=1/\left(e^{x/T}-1\right)$.

We now apply the functional renormalization group to the SU(2) linear sigma model. The core quantity in the
framework of FRG is the averaged effective action $\Gamma_k$ at the RG scale $k$ in Euclidean space, its scale dependence is described by the flow equation~\cite{Berges:2000}
\begin{equation}
\label{flow}
\partial_k\Gamma_k=\frac{1}{2}\text{Tr}\left[\frac{\partial_kR_k}{\Gamma_k^{(2)}+R_k}\right],
\end{equation}
where the trace is performed in field space and momentum space, and $\Gamma_k^{(2)}$ is the second order functional derivative of $\Gamma_k$ with respect to the field $\phi$, $\Gamma_k^{(2)}=\delta^2\Gamma_k/\delta\phi^2$. The evolution of the flow from ultraviolet limit $k=\Lambda$ to infrared limit $k=0$ encodes in principle all the quantum and thermal fluctuations in the action. To suppress the slow modes with momentum smaller than the scale $k$ during the evolution, an infrared regulator $R_k$ is introduced in the flow equation. In this work we choose the often used optimized regulator function $R_k=(k^2-\vec{p}^2)\Theta(k^2-\vec{p}^2)$~\cite{Litim:2000} at finite temperature and density. It has been demonstrated that the physics in the infrared limit is not sensitive to the choice of regularization scheme~\cite{Litim:1996}.

The averaged effective action keeps its form during the evolution, and its scale dependence is reflected in the renormalized mass $m_k$ and coupling constant $\lambda_k$. For the  linear sigma model at finite isospin chemical potential it is written as
\begin{equation}
\Gamma_k=\int d^4x\left[\frac{1}{2}\partial_\mu \pi\partial^\mu\pi+i\mu_I\left(\pi_1\partial_t\pi_2-\pi_2\partial_t\pi_1\right)+U_k(\pi)\right],
\end{equation}
where $U_k(\pi)$ is the interaction part $U(\pi)$ in (\ref{l1}) but with a replacement of $m^2$ and $\lambda$ by the renormalized $m^2_k$ and $\lambda_k$. As is well known~\cite{Berges:1998}, the contribution from the wave function renormalization to the thermodynamics of a system is much smaller in comparison with the mass and coupling constant renormalization. As a first order approximation, we have neglected here the wave function renormalization for the meson fields $\pi$.

Assuming uniform field configuration, the integral over space and imaginary time is trivial, and the effective action $\Gamma_k=\beta V U_k$ is fully controlled by $U_k$, where $V$ and $\beta=1/T$ are the space and time region of the system. Since the $O(4)$ symmetry is broken by chiral condensation and finite isospin chemical potential, the combined condensate $\langle\sigma\rangle^2+\langle\pi_1\rangle^2$ is no longer an invariance, and there are separate $\langle\sigma\rangle^2$ and $\langle\pi_1\rangle^2$ dependence in the flow equation. Making the shift $\sigma\to\langle\sigma\rangle+\sigma$ and $\pi_1\to\langle\pi_1\rangle+\pi_1$ and introducing invariance $\rho_{\sigma}=\langle\sigma\rangle^2/2$ and $\rho_{\pi}=\langle\pi_1\rangle^2/2$ and fluctuations $\delta_\sigma$ and $\delta_{\pi}$, the interaction $U_k(\pi)$ can be expressed as $U\left(\tilde{\rho}_{\sigma},\tilde{\rho}_{\pi}\right)$ with $\tilde{\rho}_{\sigma}=\rho_{\sigma}+\delta_\sigma$ and $\tilde{\rho}_{\pi}=\rho_{\pi}+\delta_\pi$. At the starting point of the evolution of the flow equation, namely at the ultraviolet scale $k=\Lambda$, the system approaches to the classical limit and all the fluctuations disappear. In this case, the interaction $U\left(\tilde{\rho}_{\sigma},\tilde{\rho}_{\pi}\right)$ is reduced to the potential $U\left(\rho_{\sigma},\rho_{\pi}\right)$. Note that, the condensates minimizing the effective potential are governed by the gap equations,
\begin{eqnarray}
&& m_k^2\langle\sigma\rangle+\lambda_k\langle\sigma\rangle^3-c=0,\nonumber\\
&& \langle\pi_1\rangle=0
\end{eqnarray}
in the normal phase and
\begin{eqnarray}
&& \langle\sigma\rangle={c\over \mu_I^2},\nonumber\\
&& \langle\pi_1\rangle=\sqrt{{\mu_I^2-m_k^2\over \lambda_k}-{c^2\over \mu_I^4}}
\end{eqnarray}
in the pion superfluidity phase. The gap equations lead to the scale dependence of the condensates $\langle\sigma\rangle_k$ and $\langle\pi\rangle_k$ and in turn the fluctuations $\delta_{\sigma k}$ and $\delta_{\pi_1 k}$ in the potential.

By taking the optimized regulator function $R_k$ and finishing the momentum integration on the righthand side of the flow equation (\ref{flow}) for $U_k\left(\tilde{\rho}_{\sigma},\tilde{\rho}_{\pi}\right)$, it can be explicitly expressed as
\begin{eqnarray}
\partial_kU_k &=&S_dk^{d+1}T\sum_n\Bigg[{1\over \omega_n^2+E_{\sigma k}^2}+{1\over \omega_n^2+E_{\pi_0 k}^2}\nonumber\\
&&+\sum_{\pm}{{1\pm\mu_I/E_{\pi_0 k}}\over \omega_n^2+\left(E_{\pi_0 k}\pm\mu_I\right)^2}\Bigg]
\end{eqnarray}
in the normal phase and
\begin{eqnarray}
\partial_kU_k &=& S_d k^{d+1} T\sum_n\left[{1\over \omega_n^2+E_{\pi_0 k}^2}+{N_k(\omega_n)\over D_k(\omega_n)}\right],\nonumber\\
N_k(\omega_n) &=&
4\mu_I^2\omega_n^2-16\lambda_k^2\tilde{\rho}_{\pi}\tilde{\rho}_{\sigma}
+\left(\omega_n^2+E_{\sigma k}^2\right)\left(\omega_n^2+E_{\pi_1 k}^2\right)\nonumber\\
&&+\left(\omega_n^2+E_{\sigma k}^2\right)\left(\omega_n^2+E_{\pi_2 k}^2\right)\nonumber\\
&&+\left(\omega_n^2+E_{\pi_1 k}^2\right)\left(\omega_n^2+E_{\pi_2 k}^2\right),\nonumber\\
D_k(\omega_n) &=& 4\mu_I^2\omega_n^2\left(\omega_n^2+E_{\sigma k}^2\right)-16\lambda_k^2\tilde{\rho}_{\pi}\tilde{\rho}_{\sigma}\left(\omega_n^2+E_{\pi_2 k}^2\right)\nonumber\\
&&+\left(\omega_n^2+E_{\sigma k}^2\right)\left(\omega_n^2+E_{\pi_1 k}^2\right)\left(\omega_n^2+E_{\pi_2 k}^2\right)
\end{eqnarray}
in the superfluidity phase, where $d$ is the dimension of the space and $S_d=1/(d~2^{d-1}\pi^{d/2}\Gamma(d/2))$ is the phase space factor, the summation is performed over the Matsubara frequency of bosons $\omega_n=2n\pi T,\ n=0,\pm 1,\pm 2,\cdots$. $E_{\phi k}=\sqrt{m_{\phi k}^2+k^2}$ are meson energies with masses $m_{\phi k}$ derived by the second order derivative of the interaction,
\begin{eqnarray}
m_{\sigma k}^2 &=& m_k^2+2\lambda_k\left(\tilde{\rho}_{\pi}+3\tilde{\rho}_{\sigma}\right),\nonumber\\
m_{\pi_1 k}^2 &=& m_k^2-\mu_I^2+2\lambda_k\left(3\tilde{\rho}_{\pi}+\tilde{\rho}_{\sigma}\right),\nonumber\\
m_{\pi_2 k}^2 &=& m_k^2-\mu_I^2+2\lambda_k\left(\tilde{\rho}_{\pi}+\tilde{\rho}_{\sigma}\right),\nonumber\\
m_{\pi_0 k}^2 &=& m_k^2+2\lambda_k\left(\tilde{\rho}_{\pi}+\tilde{\rho}_{\sigma}\right).
\end{eqnarray}
Note that, due to the spontaneous isospin symmetry breaking in the superfluidity phase, $\sigma, \pi_1$ and $\pi_2$ are not the eigenstates of the Hamiltonian of the system, the right hand side of the flow equation can no longer be expressed as a simple sum of their independent contributions.

Since $\langle\pi_1\rangle_k=0$ in the normal phase and $\langle\sigma\rangle_k=c/\mu_I^2$ in the superfluidity phase are RG scale independent, the corresponding fluctuations $\delta_{\pi k}$ in the normal phase and $\delta_{\sigma k}$ in the superfluidity phase are scale independent either. Considering the fact that any fluctuation is assumed to vanish at $k=\Lambda$, $\delta_{\pi k}$ and $\delta_{\sigma k}$ can respectively be removed from $\tilde{\rho}_{\pi}, \tilde{\rho}_{\sigma}$ and hence the the flow equation.

Expanding the both sides of the flow equation around the fluctuation $\delta_{\sigma k}$ in the normal phase and $\delta_{\pi k}$ in the superfluidity phase to the second order, and then comparing the coefficients of the linear and quadratic terms, we obtain two coupled equations for the mass parameter and coupling parameter. The derivation is straightforward, but the equations are tedious, especially in the superfluidity phase. In this way, we transferred the flow equation for the interaction $U_k$ to the flow equations for the renormalized mass parameter $m_k^2$ and coupling constant $\lambda_k$. Together with the two gap equations, their $k$-dependence are fully determined.

%%%%%%%%%%%%%%%%%%%%%%%%%%%%%%%%%%%%%%%%%%
\section{Numerical Results}
%%%%%%%%%%%%%%%%%%%%%%%%%%%%%%%%%%%%%%%%%%
In this section we numerically solve the two coupled flow equations and show our results for the phase diagram of pion superfluidity and the corresponding critical exponents at finite temperature and isospin chemical potential. The initial condition for the two flow equations at fixed temperature and chemical potential is the values of the four parameters at the ultraviolet momentum $\Lambda$, namely $m_\Lambda^2(T,\mu_I), \lambda_\Lambda(T,\mu_I), \langle\sigma\rangle_\Lambda(T,\mu_I)$ and $\langle\pi_1\rangle_\Lambda(T,\mu_I)$. Considering
the fact that the system at high enough momentum is dominated by the dynamics and not affected remarkably by the temperature and chemical potential, the temperature and chemical potential dependence of the parameters at the ultraviolet momentum can be safely neglected. Therefore, we take the temperature and chemical potential independent initial values $m_\Lambda^2(T,\mu_I)=m_\Lambda^2(0,0)$, $\lambda_\Lambda(T,\mu_I)=\lambda_\Lambda(0,0)$, $\langle\sigma\rangle_\Lambda(T,\mu_I)=\langle\sigma\rangle_\Lambda(0,0)$ and $\langle\pi_1\rangle_\Lambda(T,\mu_I)=\langle\pi_1\rangle_\Lambda(0,0)$, and they are so chosen to fit the vacuum values of meson masses $m_\pi=135$ MeV and $m_\sigma=400$ MeV and pion decay constant $f_\pi=93$ MeV at $k = 0$.

\begin{figure}[!hbt]\centering
\includegraphics[width=0.40\textwidth]{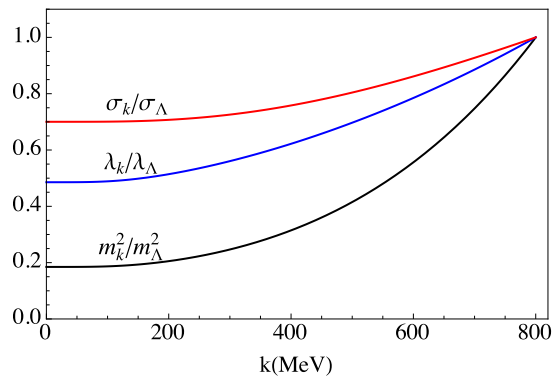}
\caption{(Color online) The evolution of the scaled mass parameter $m_k^2/m_\Lambda^2$, coupling constant $\lambda_k/\lambda_\Lambda$ and chiral condensate $\langle\sigma\rangle_k/\langle\sigma\rangle_\Lambda$ with RG scale $k$ in vacuum. }
\label{fig1}
\end{figure}

How to choose the value of the ultraviolet scale $\Lambda$ should be carefully discussed in effective models. In the spirit
of renormalization group, when more and more fluctuations
are involved in the calculation through the momentum
scale $k$ approaching from $\Lambda$ to $0$, the phase transition of a system at finite temperature and density should be reflected in the phase transition in vacuum at a critical scale $k_c$. Therefore, $\Lambda$ should be large enough to guarantee the phase transition. On the other hand, in models with hadrons as elementary constituents, the momentum scale can not go beyond the scale of the model itself where the hadrons are well defined. This means that the momentum scale
should be restricted in a reasonable region. In the following calculation we take $\Lambda=800$ MeV for the starting point of the evolution of the renormalization
parameters. We have checked that the physical results at $k=0$ are not sensitive to the further increased $\Lambda$.
In vacuum, the evolution of the three parameters $m_k^2, \lambda_k$ and $\langle\sigma\rangle_k$ with RG scale $k$ is shown in Fig.\ref{fig1}. They all drop down rapidly in the beginning and become saturated at $k\to 0$.

\begin{figure}[!hbt]\centering
\includegraphics[width=0.40\textwidth]{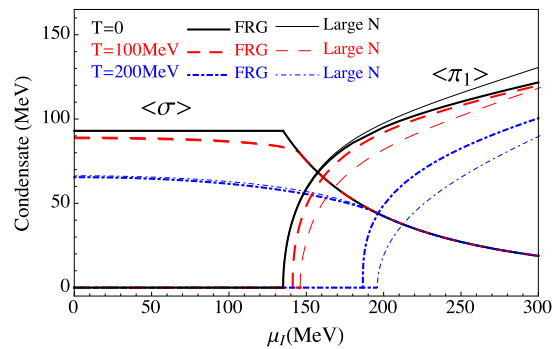}
\caption{(Color online) The chiral and pion condensates $\langle\sigma\rangle$ and $\langle\pi_1\rangle$ as functions of isospin chemical potential at different temperatures in the FRG and large $N$ expansion approaches. }
\label{fig2}
\end{figure}

The chiral and pion condensations at finite temperature and isospin chemical potential are displayed and compared with the large $N$ expansion in Fig.\ref{fig2}. The system is in normal phase at low isospin chemical potential and enters the pion superfluidity phase at the critical isospin chemical potential $\mu_I^c=m_\pi$. Due to the explicit chiral symmetry breaking ($c\neq 0$), the chiral condensate does not disappear in the superfluidity phase. While there is almost no difference between the FRG and large $N$ expansion for the chiral condensate, the difference is remarkable for the pion condensate and increases with temperature. Especially, the critical isospin chemical potential at finite temperature in the FRG is obviously lower than the value in the large $N$ expansion. This indicates that the system described by the FRG approach is easier to enter the pion superfluidity phase.

\begin{figure}[!hbt]\centering
\includegraphics[width=0.40\textwidth]{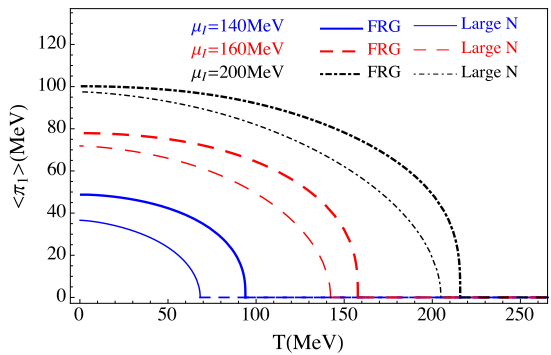}
\caption{(Color online) The pion condensate $\langle\pi_1\rangle$ as a function of temperature at different isospin chemical potential in the FRG and large $N$ expansion approaches. }
\label{fig3}
\end{figure}

The difference between the two approaches becomes more visible in Fig.\ref{fig3}. The spontaneously broken isospin symmetry in superfluidity phase can be restored by the hot medium, and the critical temperature increases with isospin chemical potential. It is clear that the pion superfluid controlled by the FRG can survive in a more hot medium in comparison with the large $N$ expansion. For instance, the critical temperature at $\mu_I=140$ MeV which is a little higher than the critical value $\mu_I^c=135$ MeV increases from 58 MeV in the large $N$ expansion to 94 MeV in the FRG.

For a continuous phase transition, while the temperature and chemical potential dependence of the order parameter and the location of the phase transition line are related to the detailed dynamics of the used model, the critical behavior of the phase transition is controlled only by the symmetry and the space-time dimension of the system. In the vicinity of a continuous phase transition, fluctuations become dominant and the correlation length approaches to infinity. In this case, the difference between the FRG which focuses on quantum and thermal fluctuations and the large $N$ expansion can be understood from the perspective of critical exponents. From the scaling laws, $\gamma=\beta(\delta-1)$, $\alpha+2\beta+\gamma=2$, $d\nu=2-\alpha$ and $\gamma=\nu(2-\eta)$, only two among the six critical exponents are independent, where $d$ is the space dimension. Since we have taken local potential approximation, the anomalous dimension disappears, $\eta=0$, and we have only one independent critical exponent.

For the phase transition of pion superfluidity, the critical exponents $\beta$ and $\nu$ describing respectively the behavior of the order parameter $\langle\pi_1\rangle$ and the singularity in the correlation length are defined as
\begin{eqnarray}
\label{exponent}
\langle\pi_1\rangle & \sim &\left({T_c-T\over T_c}\right)^\beta,\nonumber\\
m_{\pi_1} & \sim &\left({T-T_c\over T_c}\right)^\nu.
\end{eqnarray}
At zero temperature we have $d_{\text{eff}}=4$ and the scaling laws lead to $\nu=\beta$. At finite temperature and density, we can introduce an effective dimension $d_{\text{eff}}=2\beta/\nu+2$ from the scaling laws.

The external parameters like temperature and chemical potential affect not only the non-universal properties such as the location of the phase transition point but also the universal characteristics like critical exponents and universality class the system belongs to. Although the external parameters do not change the intrinsic symmetry of the system, it alters the critical exponents by reducing the space-time dimension. In finite temperature field theory~\cite{Quiros:1999,ZinnJustin:2000}, the imaginary time $it$ corresponds to the inverse temperature of the many-particle system $T^{-1}$. Therefore, the two length scales of the system defined in the Euclidean space $S^1\times R^{d}$ can be the circumference $T^{-1}$ of $S^1$ and the inverse RG-scale $k^{-1}$ for $R^d$. The space-time integration at finite temperature becomes
\begin{equation}
\int_0^\infty d^{d+1}x\to\int_0^{T^{-1}} d t\int_0^\infty d^{d}{\bf x}.
\end{equation}
In low temperature limit $T\to 0$, the integration over time is from $0$ to $\infty$, and the space-time dimension approaches to $d+1$. In high temperature limit $T\to\infty$, however, the time integration vanishes, the dimension of the system drops from $d+1$ of $S^1\times R^{d}$ to $d$ of $R^{d}$. Between the low and high temperature limits, the effective dimension of the system $d_{\text{eff}}$ changes continuously from $d+1$ to $d$.

\begin{table}[h]
\centering
\begin{tabular}{|c|c|c|c|c|c|c|c|}
\hline
$T_c\text{(MeV)}$ & 0 & 10 & 50 &100 & 150 & 200  & 250\\
\hline
$\mu_I\text{(MeV)}$ & 135 & 135.2 & 137.7&145.4 & 158.6&193.9 & 231.9\\
\hline
$\beta$ & 0.5 & 0.442 & 0.368&0.352 & 0.334&0.321 &  0.313\\
\hline
$\nu$ & 0.5 & 0.540 & 0.603&0.619 & 0.623&0.625 & 0.628 \\
\hline
$d_{\text{eff}}$ & 4 & 3.64 & 3.22&3.14 & 3.07&3.03 & 3.00 \\
\hline
\end{tabular}
\caption{The critical exponents $\beta$ and $\nu$ and effective dimension $d_{\text{eff}}$ at finite temperature and isospin chemical potential, calculated in the FRG approach.}
\label{tab1}
\end{table}

\begin{table}[h]
\centering
\begin{tabular}{|c|c|c|c|c|c|c|c|}
\hline
$T_c\text{(MeV)}$ & 0 & 10 & 50 & 100 & 150 &  200  & 250\\
\hline
$\mu_I\text{(MeV)}$ & 135 & 135.2 & 137.8 & 146.0 & 163.5 & 195.9 &242.5\\
\hline
$\beta$ & 0.5 & 0.5 & 0.5 & 0.5 & 0.5 & 0.5 & 0.5 \\
\hline
$\nu$ & 0.5 & 0.520 & 0.661 & 0.750 & 0.901 & 0.936 & 0.948 \\
\hline
$d_{\text{eff}}$ & 4 & 3.92 & 3.51 & 3.33 & 3.11 & 3.07 & 3.05 \\
\hline
\end{tabular}
\caption{The critical exponents $\beta$ and $\nu$ and effective dimension $d_{\text{eff}}$ at finite temperature and isospin chemical potential, calculated in the large $N$ expansion approach.}
\label{tab2}
\end{table}
The critical exponents $\beta$ and $\nu$ and the effective dimension $d_{\text{eff}}=2\beta/\nu+2$ at finite temperature and isospin chemical potential are shown in Table \ref{tab1} in the FRG approach and Table \ref{tab2} in the large $N$ expansion approach. From the comparison of the two approaches, the big difference lies in the critical exponents $\beta$ and $\nu$ which are controlled by fluctuations. With the large $N$ expansion, $\beta$ is temperature independent and keeps its mean field value $0.5$. In the FRG frame, however, $\beta$ decreases continuously from $0.5$ to $0.313$ with increasing temperature. In the beginning it drops down rapidly and becomes saturated when the temperature is high enough. The decreasing $\beta$ with temperature indicates that the singularity in free energy becomes more significant and thermal fluctuations are more important at higher temperature. During the decreasing process of $\beta$, the other critical exponent $\nu$ increases with temperature from the mean field value $0.5$ to $0.628$ at $T=250$ MeV. The decreasing $\beta$ and increasing $\nu$ guarantees the dropping down of the effective dimension $d_{\text{eff}}$ from $4$ at zero temperature to $3$ in high temperature limit. From zero temperature to high temperature, the system undergoes a continuous change from a 4-dimensional $O(2)$ universality class to a 3-dimensional $O(2)$ universality class. Note that the dimension reduction is a pure temperature effect. While the large $N$ expansion predicts a nonphysical constant $\beta$, the rapidly increasing $\nu$ also leads to a dimension suppression from $4$ to $3$. Of course, the suppression process is different in two approaches.

The calculation of the critical exponents does not depend on the path approaching to the phase transition point in the $T-\mu_I$ plane, since the correlation length goes to infinity
in any direction moving to the transition point. In the above calculation, $\mu_I$ is kept as a constant and the path is parallel to the $T$-axis. We checked the calculation with the path
parallel to the $\mu_I$-axis, namely taking $\langle\pi_1\rangle\sim\left(\left(\mu_I-\mu_I^c\right)/\mu_I^c\right)^\beta$ and $m_{\pi_1}\sim\left(\left(\mu_I^c-\mu_I\right)/\mu_I^c\right)^\nu$, the result is
the same as with (\ref{exponent}).

%%%%%%%%%%%%%%%%%%%%%%%%%%%%%%%%%%%%%%%%%%%%%%%%%%%%%%%%%%%%%
\section{A Compaison: O(N) Model in Continuous Dimension}
%%%%%%%%%%%%%%%%%%%%%%%%%%%%%%%%%%%%%%%%%%%%%%%%%%%%%%%%%%%%%

To show the fact that the critical exponents calculated above are independent of the details of the pion superfluidity but controlled only by the symmetry and the space-time dimension of the system, we recalculate in this section the critical exponents in a simpler model with intrinsic symmetry $O(N)$ and continuous dimension $3<d<4$. The Euclidean Lagrangian density of the model involves a set of $N$ real scalar fields $\phi_i, (i=1,\cdots,N)$,
\begin{equation}
\mathcal{L}_N=\frac{1}{2}\partial_{\mu}\phi_i\partial^{\mu}\phi_i+U(\phi^2)
\end{equation}
with the effective potential
\begin{equation}
U(\phi^2)=\frac{1}{2}a\phi_i\phi_i+\frac{1}{4}b(\phi_i\phi_i)^2,
\end{equation}
where $a$ and $b$ are respectively the mass parameter and coupling constant. Suppose one of the $N$ components is with finite vacuum expectation value $\langle\phi_j\rangle$, the Goldstone theorem guarantees $N-1$ massless particles. Defining the invariant of the system $\rho=\langle\phi_j\rangle^2/2$ and making a shift $\phi_j\to\langle\phi_j\rangle+\phi_j$, we obtain the flow equation for the effective potential,
\begin{equation}
\partial_k U_k=S_d k^{d+1}\left[\frac{1}{k^2+U'_k+2\rho U''_k}+\frac{N-1}{k^2+U'_k}\right],
\end{equation}
where $U_k(\phi^2)$ is $U(\phi^2)$ but with $k$-dependent parameters $a_k$ and $b_k$, $U'_k+2\rho U''_k$ is the squared curvature mass of $\phi_j$ with $U'_k=\partial U_k/\partial\rho$ and $U''_k=\partial^2 U_k/\partial\rho^2$. Note that the mass $U'_k$ of $N-1$ Goldstone particles is guaranteed to be zero by the gap equation.

By expanding the flow equation around the classical fields, comparing the linear and quadratic terms in fluctuations on the both sides, and taking into account the gap equation $\partial U_k/\partial\langle\phi_j\rangle=0$ which leads to only two independent parameters of $a_k, b_k$ and $\rho_k$, the flow equation for $U_k$ is converted into two coupled flow equations for $a_k$ and $b_k$. Then we introduce the dimensionless parameters,
\begin{eqnarray}
\overline a_k &=& k^{-2}a_k,\nonumber\\
\overline b_k &=& k^{d-4}b_k,
\end{eqnarray}
the flow equations for $\overline a_k$ and $\overline b_k$ can be explicitly expressed as
\begin{eqnarray}
\label{flow2}
\partial_t\overline a_t &=& -2\overline a_t
+2S_d\overline b_t\times\nonumber\\
&&\left[-\left(\frac{3}{E_1^4}+\frac{N-1}{E_2^4}\right)
+2\overline a_t\left(\frac{9}{E_1^6}+\frac{N-1}{E_2^6}\right)\right],\nonumber\\
\partial_t\overline b_t &=& (d-4)\overline b_t+4S_d\overline b_t^2\left(\frac{9}{E_1^6}+\frac{N-1}{E_2^6}\right)
\end{eqnarray}
with the RG time $t=\ln(k/\Lambda)$ and dimensionless energies $E_1=\sqrt{1-2\overline a_k}$ for the massive particle $\phi_j$ and $E_2=1$ for the $N-1$ Goldstone particles $\phi_i$. Giving an arbitrary initial condition $\overline a_0$ and $\overline b_0$ at $k=\Lambda$, we can solve the flow equations and obtain an evolution curve in the parameter plane $\overline a_t - \overline b_t$, and the collection of all the curves forms the flow diagram. From the definition, the fixed points of the flow are characterized by the equations
\begin{eqnarray}
0 &=& -2\overline a_t
+2S_d\overline b_t\times\nonumber\\
&&\left[-\left(\frac{3}{E_1^4}+\frac{N-1}{E_2^4}\right)
+2\overline a_t\left(\frac{9}{E_1^6}+\frac{N-1}{E_2^6}\right)\right],\nonumber\\
0 &=& (d-4)\overline b_t+4S_d\overline b_t^2\left(\frac{9}{E_1^6}+\frac{N-1}{E_2^6}\right).
\end{eqnarray}

\begin{figure}[!hbt]\centering
\includegraphics[width=0.40\textwidth]{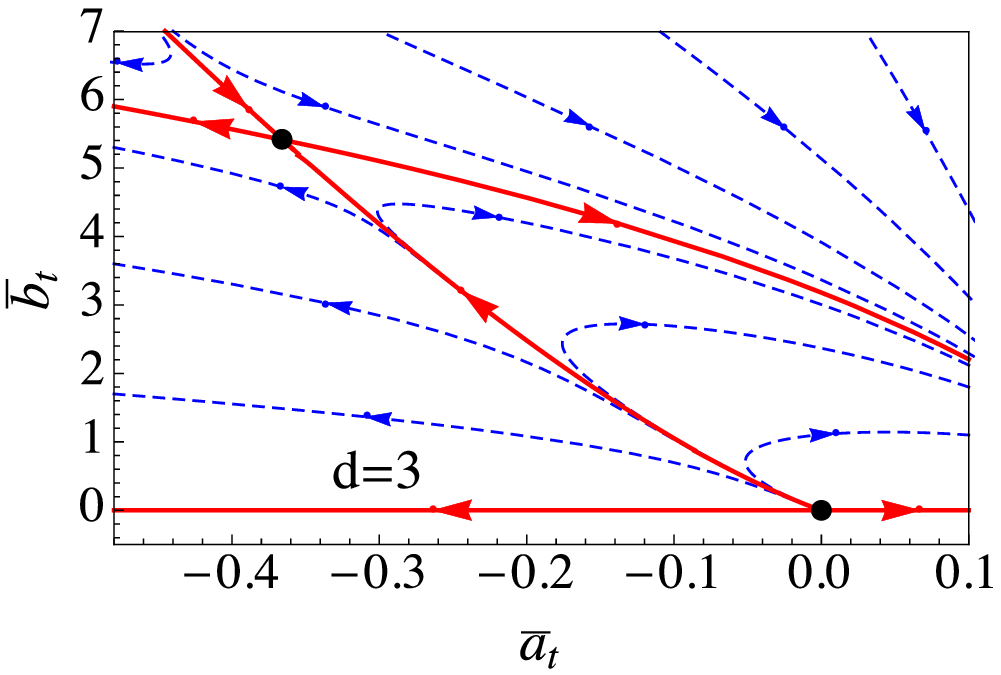}
\includegraphics[width=0.40\textwidth]{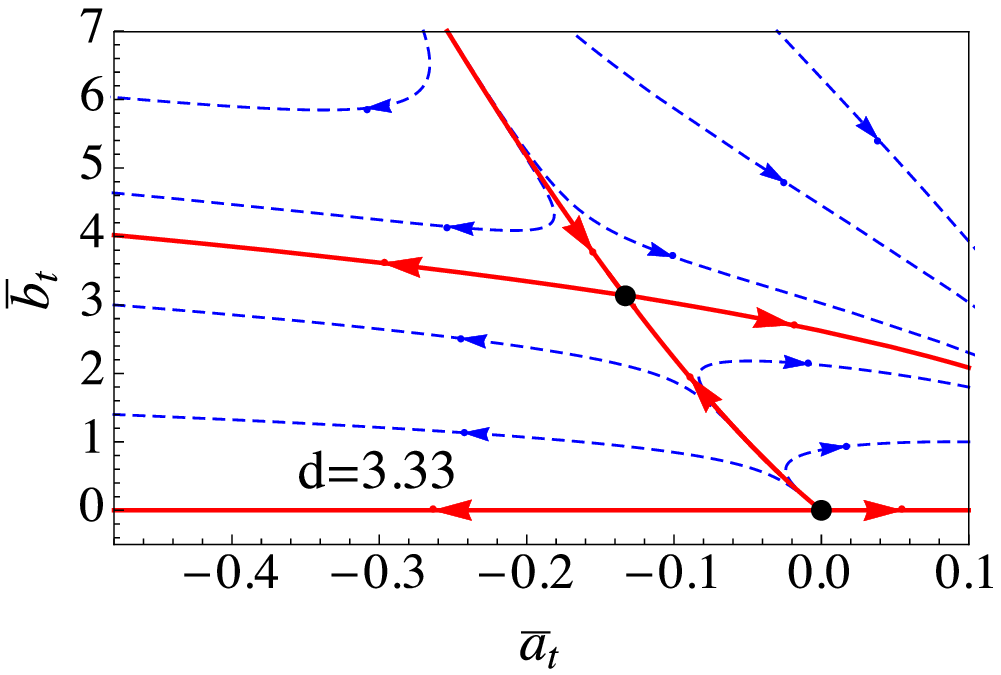}
\includegraphics[width=0.40\textwidth]{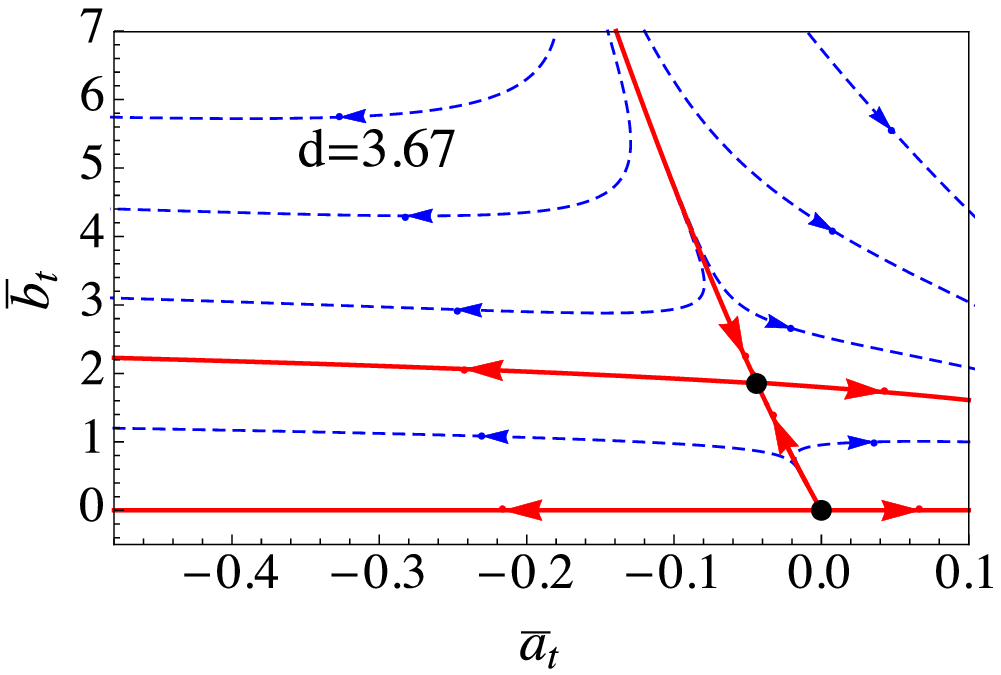}
\includegraphics[width=0.40\textwidth]{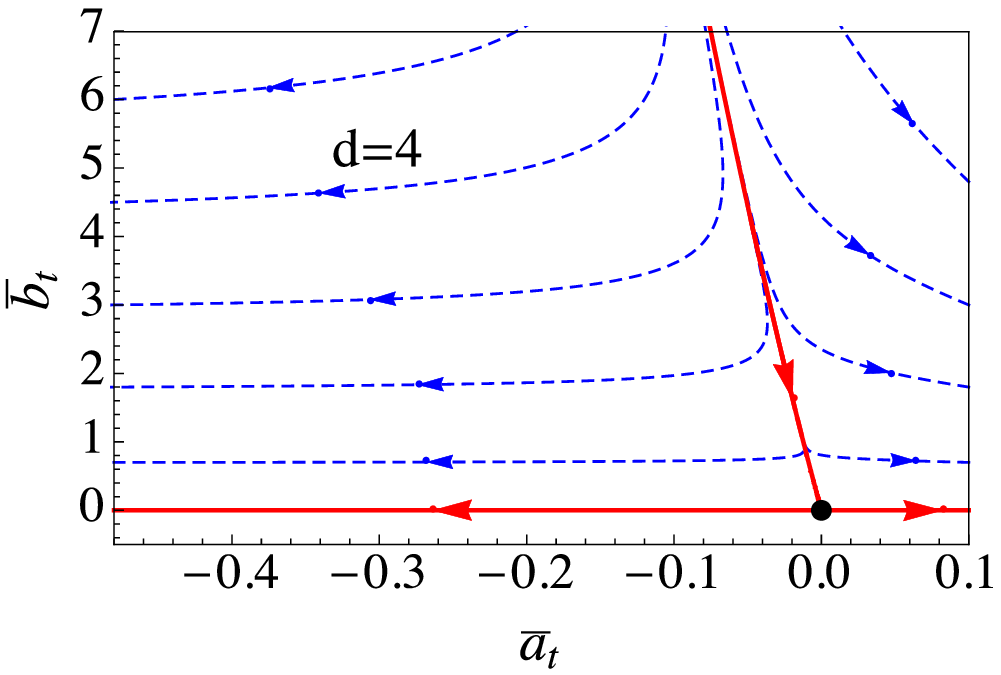}
\caption{(Color online) The flow diagrams of the $O(2)$ model in the dimensionless parameter plane $\overline a_t - \overline b_t$ at different dimensions $d=3,3{1\over 3}, 3{2\over 3}$ and $4$ from top to bottom. }
\label{fig4}
\end{figure}
Fig.\ref{fig4} shows the flow diagrams of the $O(2)$ model in the dimensionless parameter plane $\overline a_t - \overline b_t$ at different dimensions $d=3, 3{1\over 3}, 3{2\over 3}$ and $4$ from top to bottom. In the limit of $d=3$, there are two fixed points, the nontrivial one at $(\overline a_t,\overline b_t)=(-0.366,5.419)$ with flows going in and out and the trivial one at $(\overline a_t,\overline b_t)=(0,0)$ with only flows going out. The former is the Wilson-Fisher fixed point corresponding to the phase transition point in the limit of high temperature, only the flows on the critical surface indicated by thick solid lines pass through it. The other flows are attracted to go around the Wilson-Fisher fixed point but never reach it, the closer to the critical surface, the longer time the flow spends around it. The parameter $\overline a_t$ is the relevant variable, since repeated RG transformation from large $k$ to small $k$ (indicated by the arrows) drives the variable away from the fixed-point, while $\overline b_t$ is the irrelevant one, since the RG transformation pushes the flow towards the fixed point. The point at $(\overline a_t,\overline b_t)=(0,0)$ is the Gaussian fixed point, it is an ultraviolet fixed point at which the system is reduced to a free Boson system with zero rescaled coupling parameters.

When the dimension $d$ increases from $3$ to $4$, there exist both the Wilson-Fisher fixed point and Gaussian fixed point, and the critical behavior of the phase transition at finite temperature is governed by the Wilson-Fisher fixed point. During the increasing process, the Gaussian fixed point is always located at $(\overline a_t,\overline b_t)=(0,0)$, and the Wilson-Fisher fixed point approaches to it continuously and finally merges with it at $d=4$. At $d=4$, the Gaussian fixed point becomes a mixed one with flows going in and out and governs the critical behavior of the system.

With the standard procedure of renormalization group~\cite{Wilson:1971}, by linearizing the flow equations (\ref{flow2}) in the vicinity of the Wilson-Fisher fixed point (Gaussian fixed point at $d=4$) and calculating the eigenvalues of the Jacobian, we can obtain the critical exponent $\nu$ accordingly. Using the scaling law $\beta=\frac{d-2}{2}\nu$ we can further find the critical exponent $\beta$. The results for $O(N)$ models with different $N$ in continuous dimension between $3$ and $4$ are listed in Tables \ref{tab3} and \ref{tab4}.

\begin{table}[!h]
\centering
\begin{tabular}{|c|c|c|c|c|c|c|c|c|}
\hline
\backslashbox{$N$}{d} & 3 & 3.1 &  3.2 & 3.4 & 3.6 & 3.8 & 3.9 & 4\\
\hline
$O(2)$ & 0.621 & 0.596 &  0.582 & 0.563 & 0.543 & 0.521 & 0.510 & 0.5\\
\hline
$O(3)$ & 0.730 & 0.669 &  0.630 & 0.584 & 0.552 & 0.524 & 0.512 & 0.5\\
\hline
$O(4)$ & 0.796 & 0.722 &  0.668 & 0.602 & 0.560 & 0.527 & 0.513 & 0.5\\
\hline
$O(\infty)$ & 0.999 & 0.909 &  0.833 & 0.714 & 0.625 & 0.556 & 0.526 & 0.5\\
\hline
\end{tabular}
\caption{The critical exponent for order parameter $\nu$ of $O(N)$ model in continuous dimension $3\leq d\leq 4$.}
\label{tab3}
\end{table}

\begin{table}[!h]
\centering
\begin{tabular}{|c|c|c|c|c|c|c|c|c|}
\hline
\backslashbox{$N$}{d} & 3 & 3.1 &  3.2 & 3.4 & 3.6 & 3.8 & 3.9 & 4\\
\hline
$O(2)$ & 0.311 & 0.328 &  0.349 & 0.394 & 0.434 & 0.469 & 0.485 & 0.5\\
\hline
$O(3)$ & 0.365 & 0.368 &  0.378 & 0.409 & 0.442 & 0.472 & 0.486 & 0.5\\
\hline
$O(4)$ & 0.398 & 0.397 &  0.401 & 0.421 & 0.448 & 0.474 & 0.487 & 0.5\\
\hline
$O(\infty)$ & 0.5 & 0.5 &  0.5 & 0.5 & 0.5 & 0.5 & 0.5 & 0.5\\
\hline
\end{tabular}
\caption{The critical exponent for order parameter $\beta$ of $O(N)$ model in continuous dimension $3\leq d\leq 4$.}
\label{tab4}
\end{table}

At $d=4$, the critical exponents are characterized by the Gaussian fixed point of a free boson system, and the fluctuations could be neglected in the vicinity of the critical point. When the dimension is slightly less than $4$, the Wilson-Fisher fixed point is very close to the Gaussian fixed point, the system is in a weakly coupling state, and the perturbation expansion is self-consistent in this scenario. From the expansion in terms of $4-\epsilon$, the critical exponent $\nu$ for correlation length can be simplified as $\nu^{-1}=2-(N+2)/(N+8)\epsilon$ for small $\epsilon$. When $\epsilon$ is not small enough, the Wilson-Fisher point is far from the Gaussian fixed point, the system around the phase transition is in a strongly coupled state, and non-perturbative approaches are required to deal with the dominant fluctuations.

\begin{figure}[!hbt]\centering
\includegraphics[width=0.38\textwidth]{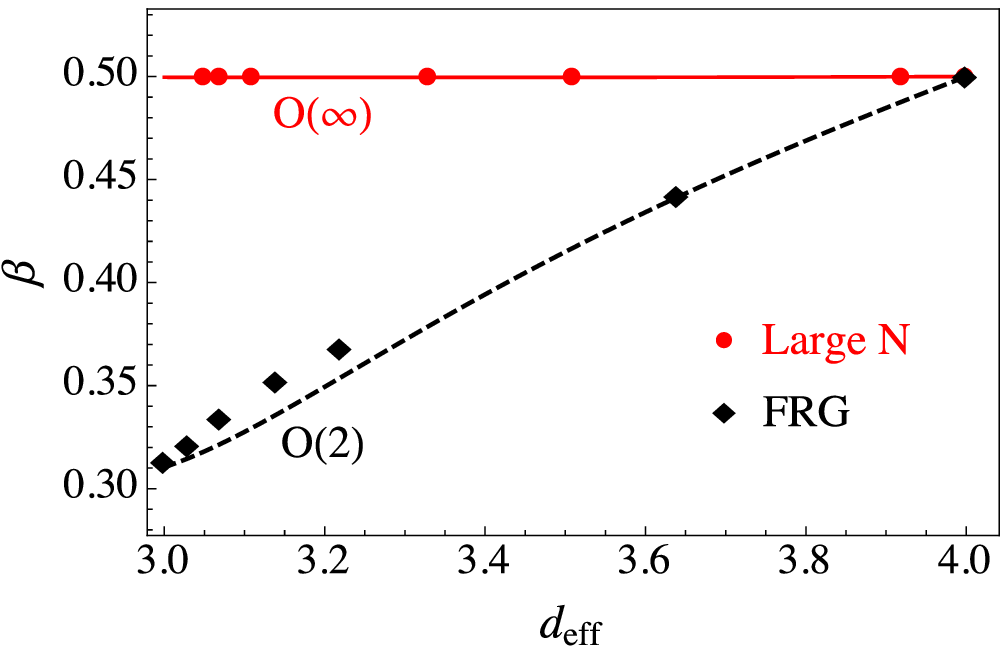}
\includegraphics[width=0.38\textwidth]{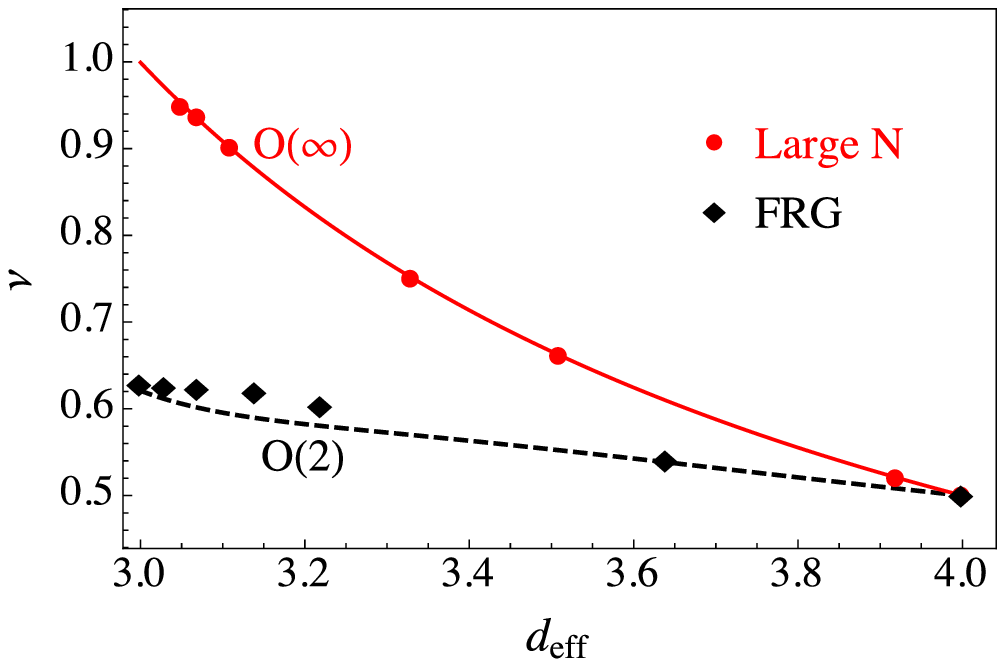}
\caption{(Color online) The critical exponents $\nu$ and $\beta$ as functions of effective dimension $d_{\text{eff}}$, calculated with general $O(N)$-type dynamics (solid lines for $O(\infty)$ and dashed lines for $O(2)$) and pion superfluidity in large $N$ expansion (dots) and FRG frame (diamonds). }
\label{fig5}
\end{figure}

We now compare the critical exponents $\nu$ and $\beta$ calculated from the fixed point analysis in the $O(N)$ model in continuous dimension (Tables \ref{tab3} and \ref{tab4}) and those calculated by linear fitting the pion condensates in FRG approach (Table \ref{tab1}) and large $N$ expansion (Table \ref{tab2}). As shown in FIG.\ref{fig5}, the large $N$ expansion (dots) is equivalent to the large $N$ limit of the $O(N)$ model in continuous dimension (solid lines), see also the last row of Tables \ref{tab3} and \ref{tab4}. The critical exponents of the pion condensate given by FRG (diamonds) agree qualitatively with the calculation of the O(2) model in dimensions between 3 and 4 (dashed lines). As we expected, the critical exponents calculated by FRG depend only on the symmetry and space-time dimension of the system and are not sensitive to the dynamical of the system: superfluidity or the general $O(2)$-type dynamics. In both cases, $\beta$ decreases form $0.5$ at $d=4$ to about $0.31$ at $d=3$, and $\nu$ increases from $0.5$ at $d=4$ to about $0.62$ at $d=3$. This indicates clearly that temperature has profound influence on the space-time dimension of the system and alters the universality class accordingly.

%%%%%%%%%%%%%%%%%%%%%%%%%
\section{Summary}
%%%%%%%%%%%%%%%%%%%%%%%%%
We studied the critical behavior of pion superfluidity at finite temperature and isospin chemical potential and its relation to the dimension crossover at zero temperature. The two relevant macroscopic parameters, temperature and  chemical potential which bring the system towards or away from the phase transition, provide us a chance to investigate the influence of them on the effective dimension near the phase transition.

In the frame of functional renormalization group, we calculated the critical exponents of the pion superfluidity in the SU(2) linear sigma model at finite temperature and isospin density and compared them with a general $O(N)$ model at zero temperature and density. The pion superfluid is a second order phase transition driven by isospin density. At zero temperature, the critical isospin chemical potential is exactly the pion mass in vacuum, and the critical exponents $\beta=0.5$ for the order parameter and $\nu=0.5$ for the correlation length belong to the $4$-dimensional $O(2)$ universality class. When temperature is turned on, the critical isospin chemical potential increases, the critical exponents $\beta$ and $\nu$ change with temperature, and the phase transition corresponds to a $O(2)$ universality class with a dimension crossover from $d_{\text{eff}}=4$ to $d_{\text{eff}}=3$. In the limit of high temperature, the critical exponents are saturated at $\beta = 0.31$ and $\nu = 0.62$, corresponding to the $3$-dimensional $O(2)$ universality class.

We also compared the FRG with large $N$ expansion in the linear sigma model. While the system undergoes a dimension crossover from $d_{\text{eff}}=4$ to $d_{\text{eff}}=3$ with increasing temperature in both approaches, the critical exponents are controlled by the way how to include fluctuations. The two approaches give the same critical behavior only at zero temperature where thermal fluctuations are negligible in the vicinity of the phase transition. This agreement is predicted by the Ginzburg criterion~\cite{Nielsen:1977} that $d=4$ is the upper critical dimension of mean field approach. From the perspective of FRG, the critical phenomenon at $d=4$ is governed by the Gaussian fixed point, around which the system is reduced to a weakly coupling one, justifying the self-consistency of mean field treatment and $4-\epsilon$ expansion. Turning on the temperature, the difference between the two approaches becomes obvious. For dimension much less than $4$, fluctuations in the critical region play the dominant role. The large $N$ expansion and perturbation treatment are no longer self-consistent, non-perturbative treatments of fluctuations are required. From the FRG point of view, the critical phenomenon at $d<4$ is governed by the Wilson-Fisher fixed point. When it is far from the Gaussian fixed point, the system is strongly coupled in the critical region.

\noindent {\bf Acknowledgement:} We thank Yin Jiang for helpful discussions during the work. The work is supported by the NSFC and MOST grant Nos. 11335005, 11575093, 2013CB922000 and 2014CB845400.

\bibliographystyle{unsrt}

\begin{thebibliography}{10}

\bibitem{Barshay:1974}
  S.~Barshay and G.~E.~Brown,
  %``Death to pion condensates in nuclear matter,''
  Phys.\ Lett.\ B {\bf 47}, 107 (1973).

 \bibitem{Khodel:2004}
  V.~A.~Khodel, J.~W.~Clark, M.~Takano and M.~V.~Zverev,
  %``Phase transitions in nucleonic matter and neutron - star cooling,''
  Phys.\ Rev.\ Lett.\  {\bf 93}, 151101 (2004).

\bibitem{Endrodi:2014}
  G.~Endr\"odi,~
  %``Magnetic structure of isospin-asymmetric QCD matter in neutron stars,''
  Phys.\ Rev.\ D {\bf 90}, no. 9, 094501 (2014).

\bibitem{Kogut:2004}
  J.~B.~Kogut and D.~K.~Sinclair,~
  %``The Finite temperature transition for 2-flavor lattice QCD at finite isospin density,''
  Phys.\ Rev.\ D {\bf 70}, 094501 (2004).

\bibitem{Kogut:2002}
  J.~B.~Kogut and D.~K.~Sinclair,~
  %``Lattice QCD at finite isospin density at zero and finite temperature,''
  Phys.\ Rev.\ D {\bf 66}, 034505 (2002).

\bibitem{He:2005}
  L.~He, M.~Jin and P.~Zhuang,~%Pion superfluidity and meson properties at finite isospin density,
  Phys.\ Rev.\ D {\bf 71}, 116001 (2005).

\bibitem{Andersen:2007}
  J.~O.~Andersen and L.~Kyllingstad,
  %``Pion Condensation in a two-flavor NJL model: the role of charge neutrality,''
  J.\ Phys.\ G {\bf 37}, 015003 (2009).

\bibitem{Abuki:2008}
  H.~Abuki, R.~Anglani, R.~Gatto, M.~Pellicoro and M.~Ruggieri,
  %``The Fate of pion condensation in quark matter: From the chiral to the real world,''
  Phys.\ Rev.\ D {\bf 79}, 034032 (2009).

\bibitem{Mu:2009}
  C.~Mu and P.~Zhuang,~%Pion Superfluidity beyond Mean Field Approximation In Nambu-Jona-Lasinio Model,
  Phys.\ Rev.\ D {\bf 79}, 094006 (2009).

\bibitem{Xia:2013}
  T.~Xia, L.~He and P.~Zhuang,~%Three-flavor Nambu-Jona-Lasinio model at finite isospin chemical potential,
  Phys.\ Rev.\ D {\bf 88}, 056013 (2013).

\bibitem{Kamikado:2012}
  K.~Kamikado, N.~Strodthoff, L.~von Smekal and J.~Wambach,~%Fluctuations in the quark-meson model for QCD with isospin chemical potential,
  Phys.\ Lett.\ B {\bf 718}, 1044 (2013).

\bibitem{Svanes:2010}
  E.~E.~Svanes and J.~O.~Andersen,~%Functional renormalization group at finite density and Bose condensation,
  Nucl.\ Phys.\ A {\bf 857}, 16 (2011).

\bibitem{Phat:2011}
  T.~H.~Phat and N.~V.~Thu,
  %``Phase structure of the linear sigma model with the non-standard symmetry breaking term,''
  J.\ Phys.\ G {\bf 38}, 045002 (2011).

\bibitem{Floerchinger:2009}
  S.~Floerchinger, R.~Schmidt, S.~Moroz and C.~Wetterich,
  %``Functional renormalization for trion formation in ultracold fermion gases,''
  Phys.\ Rev.\ A {\bf 79}, 013603 (2009).

\bibitem{Boettcher:2012}
  I.~Boettcher, J.~M.~Pawlowski and S.~Diehl,
  %``Ultracold atoms and the Functional Renormalization Group,''
  Nucl.\ Phys.\ Proc.\ Suppl.\  {\bf 228}, 63 (2012).

\bibitem{Friman:2012}
  B.~Friman, K.~Hebeler and A.~Schwenk,
  %``Renormalization group and Fermi liquid theory for many-nucleon systems,''
  Lect.\ Notes Phys.\  {\bf 852}, 245 (2012).

\bibitem{Drews:2013}
  M.~Drews, T.~Hell, B.~Klein and W.~Weise,
  %``Thermodynamic phases and mesonic fluctuations in a chiral nucleon-meson model,''
  Phys.\ Rev.\ D {\bf 88}, no. 9, 096011 (2013).

\bibitem{Schaefer:1999}
  B.~J.~Schaefer and H.~J.~Pirner,
  %``Renormalization group flow and equation of state of quarks and mesons,''
  Nucl.\ Phys.\ A {\bf 660}, 439 (1999).

\bibitem{Bohr:2000}
  O.~Bohr, B.~J.~Schaefer and J.~Wambach,
  %``Renormalization group flow equations and the phase transition in O(N) models,''
  Int.\ J.\ Mod.\ Phys.\ A {\bf 16}, 3823 (2001).

\bibitem{Blaizot:2006}
  J.~P.~Blaizot, A.~Ipp, R.~Mendez-Galain and N.~Wschebor,
  %``Perturbation theory and non-perturbative renormalization flow in scalar field theory at finite temperature,''
  Nucl.\ Phys.\ A {\bf 784}, 376 (2007).

\bibitem{Braun:2008}
  J.~Braun,
  %``The QCD Phase Boundary from Quark-Gluon Dynamics,''
  Eur.\ Phys.\ J.\ C {\bf 64}, 459 (2009).

\bibitem{Braun:2009}
  J.~Braun, L.~M.~Haas, F.~Marhauser and J.~M.~Pawlowski,
  %``Phase Structure of Two-Flavor QCD at Finite Chemical Potential,''
  Phys.\ Rev.\ Lett.\  {\bf 106}, 022002 (2011).

\bibitem{Stokic:2009}
  B.~Stokic, B.~Friman and K.~Redlich,
  %``The Functional Renormalization Group and O(4) scaling,''
  Eur.\ Phys.\ J.\ C {\bf 67}, 425 (2010).

\bibitem{Fukushima:2010}
  K.~Fukushima, K.~Kamikado and B.~Klein,
  %``Second-order and Fluctuation-induced First-order Phase Transitions with Functional Renormalization Group Equations,''
  Phys.\ Rev.\ D {\bf 83}, 116005 (2011).

\bibitem{Herbst:2013}
  T.~K.~Herbst, J.~M.~Pawlowski and B.~J.~Schaefer,
  %``Phase structure and thermodynamics of QCD,''
  Phys.\ Rev.\ D {\bf 88}, no. 1, 014007 (2013).

\bibitem{Appelquist:1981}
  T.~Appelquist and R.~D.~Pisarski,
  %``High-Temperature Yang-Mills Theories and Three-Dimensional Quantum Chromodynamics,''
  Phys.\ Rev.\ D {\bf 23}, 2305 (1981).

\bibitem{Liao:1996}
  S.~B.~Liao and M.~Strickland,
  %``Dimensional crossover and effective exponents,''
  Nucl.\ Phys.\ B {\bf 497}, 611 (1997).

\bibitem{Ballhausen:2003}
  H.~Ballhausen, J.~Berges and C.~Wetterich,
  %``Critical phenomena in continuous dimension,''
  Phys.\ Lett.\ B {\bf 582}, 144 (2004).

\bibitem{Nielsen:1977}
J.~Als-Nielsen and R.~J.~Birgeneau,~
%``Mean field theory, the Ginzburg criterion, and marginal dimensionality of phase,''
Am.\ J. \ Phys.{\bf 45}, 554 (1977).

\bibitem{Wilson:1971}
  K.~G.~Wilson and M.~E.~Fisher,
  %``Critical exponents in 3.99 dimensions,''
  Phys.\ Rev.\ Lett.\  {\bf 28}, 240 (1972).

\bibitem{Guida:1998}
  R.~Guida and J.~Zinn-Justin,
  %``Critical exponents of the N vector model,''
  J.\ Phys.\ A {\bf 31}, 8103 (1998).

\bibitem{Nemoto:1999}
  Y.~Nemoto, K.~Naito and M.~Oka,
  %``Effective potential of O(N) linear sigma model at finite temperature,''
  Eur.\ Phys.\ J.\ A {\bf 9}, 245 (2000).

\bibitem{Baacke:2002}
  J.~Baacke and S.~Michalski,
  %``The O(N) linear sigma model at finite temperature beyond the Hartree approximation,''
  Phys.\ Rev.\ D {\bf 67}, 085006 (2003).

\bibitem{Andersen:2008}
  J.~O.~Andersen and T.~Brauner,
  %``Linear sigma model at finite density in the 1/N expansion to next-to-leading order,''
  Phys.\ Rev.\ D {\bf 78}, 014030 (2008).

\bibitem{Andersen:2004}
  J.~O.~Andersen, D.~Boer and H.~J.~Warringa,
  %``Thermodynamics of O(N) sigma models: 1/N corrections,''
  Phys.\ Rev.\ D {\bf 70}, 116007 (2004).

\bibitem{Petropoulos:2004}
  N.~Petropoulos,
  %``Linear sigma model at finite temperature,''
  hep-ph/0402136.

\bibitem{Moshe:2003}
  M.~Moshe and J.~Zinn-Justin,
  %``Quantum field theory in the large N limit: A Review,''
  Phys.\ Rept.\  {\bf 385}, 69 (2003).

\bibitem{Haber:1981}
H.~E.~Haber and H.~A.~Weldon,
  %``Finite Temperature Symmetry Breaking as Bose-Einstein Condensation,''
  Phys.\ Rev.\ D {\bf 25}, 502 (1982).

\bibitem{Berges:2000}
  J.~Berges, N.~Tetradis and C.~Wetterich,~%``Nonperturbative renormalization flow in quantum field theory and statistical physics,''
  Phys.\ Rept.\  {\bf 363}, 223 (2002).

\bibitem{Litim:2000}
  D.~F.~Litim,~%Optimization of the exact renormalization group,
  Phys.\ Lett.\ B {\bf 486}, 92 (2000).

\bibitem{Litim:1996}
  D.~F.~Litim,
  %``Scheme independence at first order phase transitions and the renormalization group,''
  Phys.\ Lett.\ B {\bf 393}, 103 (1997).

\bibitem{Berges:1998}
  J.~Berges, D.~U.~Jungnickel and C.~Wetterich,
  %``The Chiral phase transition at high baryon density from nonperturbative flow equations,''
  Eur.\ Phys.\ J.\ C {\bf 13}, 323 (2000).

\bibitem{Quiros:1999}
  M.~Quiros,
  %``Finite temperature field theory and phase transitions,''
  hep-ph/9901312.

\bibitem{ZinnJustin:2000}
  J.~Zinn-Justin,
  %``Quantum field theory at finite temperature: An Introduction,''
  hep-ph/0005272.


\end{thebibliography}

\end{document}